# Surface symmetry breaking and disorder effects on superconductivity in perovskite BaBi$_3$ epitaxial films

Wen-Lin Wang[1], Yi-Min Zhang[1], Nan-Nan Luo[1], Jia-Qi Fan[1], Chong Liu[1], Zi-Yuan Dou[1], Lili Wang[1,2], Wei Li[1,2], Ke He[1,2], Can-Li Song[1,2,*], Yong Xu[1,2,3,†], Wenhui Duan[1,2,4], Xu-Cun Ma[1,2], Qi-Kun Xue[1,2,‡]

[1]*State Key Laboratory of Low-Dimensional Quantum Physics, Department of Physics, Tsinghua University, Beijing 100084, China*

[2]*Collaborative Innovation Center of Quantum Matter, Beijing 100084, China*

[3]*RIKEN Center for Emergent Matter Science (CEMS), Wako, Saitama 351-0198, Japan*

[4]*Institute for Advanced Study, Tsinghua University, Beijing 100084, China*

**The structural or electronic symmetry breaking of the host lattice is a recurrent phenomenon in many quantum materials, including superconductors. Yet, how these broken symmetry states affect the electronic pair wave function of superconductivity have been rarely elucidated. Here, using low-temperature scanning tunneling microscopy and first-principles calculations, we identify the broken rotational symmetry via stripe ordering on the (001) surface of perovskite BaBi$_3$ films grown by molecular beam epitaxy, and show that it consequently leads to anisotropic superconductivity with twofold symmetry. In contrast, the structural disorder smears out the anisotropy of electron pairing and fills superconducting subgap density of states as the film thickness is reduced. A quasi-long range model of superconducting fluctuations is revealed to describe the tunneling conductance spectra of thin BaBi$_3$ films well, and to exemplify how disorders contribute to the low-energy quasiparticle excitations in superconductors. Our findings help understand the effects of symmetry breaking states and disorders on superconductivity, particularly the existing tunneling conductance spectra there.**

*Correspondence and requests for materials should be addressed to*

*clsong07@mail.tsinghua.edu.cn, yongxu@mail.tsinghua.edu.cn, qkxue@mail.tsinghua.edu.cn.*

## I. INTRODUCTION

Superconductivity often emerges in proximity to and even co-exists with a plethora of broken symmetry states, e.g. charge/spin density waves [1-5], structural distortion [6, 7], electronic nematicity [8-14], static and fluctuating stripes [15-18] etc. These modulated states are inextricably intertwined with superconductivity and lead to very sophisticated phase diagrams, irrespective of whether the nature of superconductivity is conventional or not [19, 20]. Recent experimental and theoretical advances have been establishing $C_2$-symmetric stripe phases as universal in these materials [18, 21-25], which are deemed crucial for the formation of Cooper pairs. However, the subtle interrelation between the intriguing phases and superconductivity − whether they compete [19, 21, 26], correlate with [8, 9, 18, 22-25], or how they act on the emergent superconducting pairing [27-30] − remains a hotly debated topic. On the other hand, the occurrence of superconductivity relies mostly on chemical dopants or interface engineering, which turn out to be potential sources of disorder to intertwine with the existing broken symmetry states and superconductivity [31-35], masking many intrinsic effects. For superconductors with such a variety of interactions, tracing the respective roles of symmetry breaking and disorder played in superconductivity remains essential but challenging in experiment. It necessaries the exploration of simple superconducting systems where the symmetry breaking and disorder effects can be easily controlled.

In the present study, we introduce a non-oxide perovskite superconductor $BaBi_3$ with tetragonal crystal structure of space group P4/*mmm* as shown in Fig. 1(a). Its bulk becomes superconducting below a critical temperature $T_c$ = 5.9 K [36], recently being studied for pursuing novel electronic states related to the strong spin-orbit coupling of Bi [36-38]. Remarkably, $BaBi_3$ has a much simpler structure as compared with other perovskite oxide superconductors (e.g. cuprates [3, 5, 7-9, 16, 21, 22, 26] and bismuthates [18]), and offers an ideal platform to experimentally elucidate the effects of symmetry breaking and disorder on the electronic pair wave function of superconductivity. We grow $BaBi_3$ epitaxial films with fine control of film thickness on $SrTiO_3$(001) substrates using state-of-the-art molecular beam epitaxy (MBE) technique. Our scanning tunneling microscopy (STM) study reveals a series of stripe-like structures on the surface of $BaBi_3$ epitaxial films, which breaks the $C_4$ rotational symmetry of the host crystal, and also structural disorder induced by the $SrTiO_3$(001) substrates, which enhances in strength with reducing film thickness. Tunneling experiments discover twofold anisotropic superconducting gap function by surface symmetry breaking in the thick films, but reveal in the thin films that disorder smears out the gap anisotropy, fills the subgap density of states (DOS) and suppresses superconducting coherence peaks. We find that the resultant "soft" spectral gaps in disordered thin films are all satisfactorily modelled in the framework of quasi-long range superconducting fluctuations.

## II. METHODS

All the experiments were performed in an ultrahigh vacuum (UHV) cryogenic STM apparatus equipped with a MBE system for *in-situ* sample preparation (Unisoku). The base pressure for both systems is better than $< 1.0 \times 10^{-10}$ Torr. Nb-doped SrTiO$_3$(001) substrates (0.05 wt%) were degassed at 600°C for 3 h, and then annealed at 1250°C for 20 min to obtain clean surface. The samples studied, namely Ba(Sr)Bi$_3$ crystalline films, were epitaxially grown on SrTiO$_3$ substrate by co-evaporating high-purity Ba(Sr) (99.9%) and Bi (99.997%) sources from their respective Knudsen cells, with the substrate held at 340°C (240°C). The growth rate was approximately 0.1 unit cell per minute. By varying the growth duration, the film thickness can be controlled as wanted. After MBE growth, the samples were *in-situ* transferred from MBE chamber to STM stage for topographic and spectroscopic measurements. Prior to data collection, polycrystalline PtIr tip was cleaned by electron-beam bombardment and appropriately calibrated on Ag/Si(111) films. Tunneling conductance d$I$/d$V$ spectra and ZBC maps were acquired at 0.4 K by disrupting the feedback circuit, sweeping the sample voltage $V$, and recording the differential conductance d$I$/d$V$ value using a standard lock-in technique with a small bias modulation of 0.1 mV at 913 Hz, unless other specified. The tunneling gap for all d$I$/d$V$ spectra and maps is set at $V = 10$ mV and $I = 200$ pA, unless otherwise specified.

First-principles calculations were performed within the framework of density functional theory (DFT), as implemented in the Vienna ab initio simulation package[37]. The Perdew-Burke-Ernzerhof exchange correlation[38], the projector-augmented-wave potential[39] and the plane-wave basis with an energy cutoff of 400 eV were employed. The period slab approach was applied to model the BaBi$_3$ and SrBi$_3$ surfaces, which utilized a six-layer slab with the bottom two layers fixed during structural relaxation and a vacuum layer of ~ 13 Å. Structural optimization was performed using a force convergence criterion of 0.01 eV/Å. $12 \times 12 \times 12$, $10 \times 10 \times 1$ and $5 \times 5 \times 1$ Monkhorst-Pack k grids together with a Gaussian smearing of 0.05 eV were used in self-consistent calculations of bulk, $\sqrt{2} \times \sqrt{2}$ and $2\sqrt{2} \times 2\sqrt{2}$ surface supercells, respectively. The spin-orbit coupling was included in all the calculations except for structural relaxation. STM images were simulated by the local DOS based on the Tersoff-Hamann approach[40]. The surface formation energy was calculated as a function of atomic chemical potential $\mu_i$ for a $\sqrt{2} \times \sqrt{2}$ surface supercell that has $n_i$ atoms of the $i_{th}$ species more than the reference surface (i.e. the ideal Ba-Bi or Sr-Bi terminated surface): $\Delta E = E - E_0 - \sum_i n_i \mu_i$, where $E$ and $E_0$ are the total energies of the given and reference surfaces, respectively. The calculated bulk lattice constants are $a = b = 5.174$ Å and $c = 5.130$ Å for $BaBi_3$ and $a = b = c = 5.081$ Å for $SrBi_3$.

## III. RESULTS AND DISCUSSIONS

### A. MBE growth and surface structure characterization

Figure 1(b) depicts a constant-current STM topographic image of as-prepared $BaBi_3$ films, which straddle continuously over the underlying $SrTiO_3$ step edges. This bears a similarity with that of MBE-grown β-$Bi_2$Pd films [43]. Shown in Fig. 1(c) is a magnified STM image of $BaBi_3$ with a square lattice structure, presenting a preferential growth of epitaxial films along the (001) direction. This is unambiguously confirmed by analysis of the Moiré superstructure of $BaBi_3$ films on $SrTiO_3$ [Figs. S1 and S2], as explained in the Supplementary Material [44]. The spacing of neighboring bright spots is measured to be approximately 7.4 Å, which is about $\sqrt{2}$ times of the lattice constant $a = 5.188$ Å of $BaBi_3$ in the $a$-$b$ plane [37]. This means a $\sqrt{2} \times \sqrt{2}$ reconstructed structure formed on the surface. Notably point defects are observed to occupy the hollow sites of the square spots [Fig. 1(d)], as a higher flux ratio of Bi/Ba is used during the film growth.

In addition to the $\sqrt{2} \times \sqrt{2}$ surface reconstruction, another prominent feature of the surface is the occurrence of various stripe-like patterns [Figs. 1(c) and S3], which run along either the [1 0 0] or [0 1 0] azimuths ($a$ or $b$). These stripe patterns are differently spaced and have a dominant separation of 6$a$ [Fig. S3], resembling a devil's staircase behavior. The stripe height is measured to be 0.1 Å at 5 mV. It should be emphasized that both the $\sqrt{2} \times \sqrt{2}$ reconstruction and stripes (which always run along either the $a$ or $b$ axes) are universally recognized on $BaBi_3$ regardless of the film thickness, although the fourfold periodic ripples from Moiré patterns triumph over the stripe corrugation and render them not easily visible in ultrathin films [Fig. 1(e)]. Here the Moiré superstructure-induced ripples distort the crystal lattice and serve as potential scatters to influence the superconducting pairing of BaBi$_3$ [29-32]. Thus far, the stripe phases, breaking the $C_4$ rotational symmetry, in conjunction with the tunable structural disorder from the Moiré ripples by varying the film thickness, make BaBi$_3$ a promising research platform for exploring the interplay between symmetry breaking, disorder and superconductivity.

### B. Gap spectroscopy and vortex structure in the superconducting state

In order to find the effect of symmetry breaking on superconductivity, we firstly measured the differential conductance d$I$/d$V$ spectra on 34 unit cell (UC) thick BaBi$_3$ epitaxial film with small Moiré ripple. Here the film thickness is determined by measuring its height from the substrate and dividing it by the out-of-plane lattice constant of 5.130 Å, owing to the island growth mode of BaBi$_3$ on SrTiO$_3$ [Fig. S1(a)]. The tunneling spectra are typified in Fig. S4, where a superconducting gap near the Fermi level ($E_F$) is clearly visible. Plotted in Fig. 2(a) are a series of superconducting gaps acquired on a region of 272 nm × 272 nm at 0.4 K. The tunneling gaps all exhibit vanishing spectral weight over a finite energy range near $E_F$ and two clear coherence peaks at the gap edges. These features, characteristics of full-gap superconductivity [35], provide the first tunneling evidence of no gap node in the superconducting pairing function of BaBi$_3$, as anticipated. To compare the experimental curves with theoretical ones, we normalize the tunneling conductance by dividing the spatially averaged d$I$/d$V$ spectrum by its background (that is extracted from a linear fit to the conductance for |$V$| > 3.5 mV) and illustrate it in Fig. 2(b) as black symbols. In theory, such spectrum can be described by the BCS Dynes formula with an effective energy broadening Γ [45],

$$\frac{dI}{dV}(V) \cong \mathrm{Re}\left[\frac{V - i\Gamma}{\sqrt{(V - i\Gamma)^2 - \Delta^2}}\right]. \quad (1)$$

Attempts to fit the experimental data in Fig. 2(b) with a BCS $s$-wave isotropic gap Δ based on Eq. (1) are never successful. The best fitting (magenta line), which not yet explains well the local DOS near $E_F$, yields a gap magnitude Δ of 1.06 meV and unreasonably large Γ of 0.15 meV. Although a smaller Γ will remove the spectral weight near $E_F$ and lead to the experimentally observed flat bottom, it gives rise to very sharp coherence peaks, making the global fitting even worse.

The situation is exacerbated as the tunneling spectra of thinner BaBi$_3$ films are concerned. Figure 2(c) represents eight d$I$/d$V$ spectra along a 73-nm trajectory on 5 UC thick film. As compared to those on thick films [Figs. 2(a) and 2(b)], the coherence peaks are more appreciably suppressed and the subgap DOS near $E_F$ is significantly lifted. The resultant V-shaped "soft" gaps lie as well beyond the conventional BCS *s*-wave prediction, as exemplified in Fig. 2(d).

Vortex imaging of type-II superconductors provides unique capability to bring insight into superconducting gap structure. Figures 3(a-c) display spatial-resolved zero-bias conductance (ZBC) maps on $BaBi_3$ epitaxial films of varying thickness, with the magnetic fields applied perpendicular to the *ab*-plane of $BaBi_3$. The superconducting order parameter vanishes within the vortex cores, allowing ZBC maps to reveal vortex cores as yellow regions with enhanced ZBC. Intriguingly, their shapes are dependent on the thickness of $BaBi_3$ films, and exhibit a gradual crossover from elliptical [Fig. 3(a)] to circular [Fig. 2(c)] as the film thickness decreases. A close scrutiny of the elliptical vortices and the corresponding STM topographies reveals that they are always elongated along the surface stripes [cf. Fig. 3(d) and Fig. 3(e) or Fig. 3(b) and Fig. S5]. It is worth noticing that two orthogonal stripes could intersect to form grain boundaries [Figs. S3(f) and S3(g)], across which the long and short axes of elongated vortex cores interchange [Fig. 3(b)]. These observations compellingly establish a close link between the vortex core structure and surface stripes.

Previously, the elongated vortices have been observed only rarely in cuprate superconductor $YBa_2Cu_3O_{7-\delta}$ with *ab*-plane anisotropy ($a \neq b$) [46], FeSe with strong electronic nematicity [28], and quasi-one-dimensional compound $Ta_4Pd_3Te_{16}$ [47, 48] and Ni-Bi compounds [49]. In principle, the elongated vortex cores can be readily accounted for by the anisotropy of superconducting coherence length $\xi$, which scales with Fermi velocity $v_F$ and $1/\Delta$ via $\xi = \hbar v_F/\pi\Delta$. knowing that the surface stripe phases work as the only source for breaking the fourfold symmetry of $BaBi_3$, we claim that it is the surface stripe phase to generate the twofold anisotropy in $v_F$ and/or $\Delta$, and consequently lead to the elongated vortices on thick films. On the other hand, the circular vortex cores on thin $BaBi_3$ films, despite the preservation of surface stripe, implies that other factors are involved to kill the anisotropy in $v_F$ and/or $\Delta$. Since we prepared the $BaBi_3$ films using the same recipe, the observed differences between thick and thin films are probably caused by a strength distinction of Moiré ripples. A straightforward explanation would be that the Moiré ripples cause strong electron scattering and smear out the gap anisotropy of $\Delta$ in thin films, considering that disorder scattering typically has minor influence on the band structure and the corresponding $v_F$. In other words, it is the surface stripe phase that breaks the $C_4$ symmetry and leads to a twofold symmetric superconducting gap $\Delta$ and elongated vortices on thick films, while the structural disorders from Moiré ripples tend to kill the gap anisotropy and result in the circular-shaped vortex cores on thin films. Indeed, by using a twofold symmetric gap function $\Delta(\theta) = \Delta_1 + \Delta_2\cos(2\theta)$, the theoretical curve (blue line in Fig. 2(b)) well reproduces the results of thick BaBi$_3$ films, yielding $\Delta_1 = 1.01$ meV and $\Delta_2 = 0.28$ meV. The quantitative agreements, both at the gap bottom and near the coherence peaks, confirm the twofold symmetric gap function on thick BaBi$_3$ films.

To quantify the elongation of vortex cores on 34 UC thick film (the thickest one we studied), the radial dependences of vortex-induced ZBC variations acquired along and perpendicular to the stripes are plotted in Fig. 3(f). Based on the Ginzburg-Landau description of the order parameter near a superconductor-metal interface [47, 48], the ZBC profile across the vortex core obeys $Z(r) = Z_\infty + (1 - Z_\infty)[1 - \tanh(-r/\sqrt{2}\xi)]$, where $Z(r)$ represents the normalized ZBC at the distance $r$ to the vortex core. The best fits to our experimental data yield the superconducting coherence length $\xi_a = 12.4$ nm, $\xi_b = 17.3$ nm, and the anisotropy $\xi_b/\xi_a = 1.4$. The averaged coherence length $\bar{\xi} = \sqrt{\xi_a \xi_b} = 14.6$ nm is close to the reported values of bulk BaBi$_3$ superconductor [36, 38]. This value appears several times larger than the separations between adjacent stripe patterns [Figs. 1(c) and S3]. It is therefore anticipated that the surface stripes would lead to an anisotropic pair potential and order parameter $\Delta$ [Fig. 2(b)], but little alter the *dI/dV* spectra in the microscopic scale [47, 48]. Otherwise, the SC correlation function, electronic DOS and the superconductivity will be localized like nanoscale stripe structure, and spatially inhomogeneous superconducting gaps might be observed across the stripes.

The twofold symmetric pairing in thick BaBi$_3$ films is further evidenced by studying the spatial evolution of d*I*/d*V* spectra in the vicinity of a single vortex, as shown in Fig. 3(g). At the very center of vortex core, a prominent ZBC peak, which we attribute to the Andreev bound states originating from the constructive interference between electron-like and hole-like quasiparticles within vortices [50], is clearly identified. The ZBC peaks decay outside the vortex core, but behave differently along the *a* and *b* axes. Although the bound states along the *a* axis completely vanish, they keep discernable along the *b* axis, at a distance $r = 8.4$ nm from the vortex center. No splitting of the ZBC peak is resolvable when measured away from the vortex center [Fig. 3(g)]. This is possibly caused by the small energy level spacing $\varepsilon_0 = \Delta^2/E_F$ of the discrete vortex bound states [44], which is exceeded by the thermal broadening of 0.4 K, but also probably by the disorder-induced intrinsic broadening of the quasiparticle states. The latter is actually implied by exploring the vortex core excitations of thin films. As displayed in Fig. 3(h), no ZBC peak can be observed at the vortex center of 5 UC films, where the enhanced disorder scattering significantly reduces the electron mean free path $\ell$ and pushes the thin films to the dirty limit ($\ell < \xi$, no constructive interference of quasiparticles and thus no ZBC peak). Here the transition of BaBi$_3$ films from the clean ($\ell > \xi$) to dirty limit bears a strong resemblance to the previously observed ones in 2*H*-Nb$_{1-x}$Ta$_x$Se$_2$ and Pb [51, 52], triggered by the chemical [51] and interfacial disorder [32], respectively.

Notably, the twofold symmetric gap observed on thick BaBi$_3$ films [Fig. 2(b)] does not mean that the bulk BaBi$_3$ is characteristic of anisotropic order parameter Δ. Instead, the bulk superconducting gap Δ should be isotropic [34], since the stripes occur primarily at the top surface and no symmetry breaking is involved in bulk. Considering the STM sensitivity to surface DOS, the measured *dI/dV* spectra have their origin from the anisotropic surface order parameter Δ caused by the stripe phases there, leaving the bulk DOS inaccessible. As the magnetic vortices penetrate through the surface-near regions, the internal cores are subject to the anisotropic pair potential and will be elliptically deformed for energetically favorable state, just as those of anisotropic superconductors [47, 48].

The disorder response of vortex configuration has also been studied by mapping the spatial ZBC on a larger field of view of 454 nm × 454 nm, with the same pixels of 64 × 64. To better visualize the vortices, we find every vortex center by the local ZBC maximum and draw the Voronoi cells on the ZBC maps in Figs. 4(a) and 4(b). From the calculated Voronoi cell size, we estimate the average flux per vortex $\Phi_0 = 2.05 \times 10^{-15}$ Wb, consistent with a single flux quantum of $2.07 \times 10^{-15}$ Wb. Evidently, vortices do not arrange into the ordered hexagonal or square lattice. Instead, a distorted hexagonal lattice is justified from the autocorrelation image of vortex lattice, as illustrated in Fig. 4(c). This underscores the role of vortex pinning played primarily by the structural disorders of epitaxial BaBi$_3$ films. Figure 4(d) shows the Delaunay triangulation analysis of Fig. 4(b), in which every vertex denotes the position of vortex and is color-coded by its coordination number. Only half of vortices are six-fold coordinated, affirming the occurrence of vortex pinning in BaBi$_3$ films.

We further evaluate the vortex pinning by measuring the relative distances $d_{ij}=|r_i - r_j|$ for all vortex pairs at positions $r_i$ and $r_j$. The histograms of such distances are plotted in Fig. 4(e), which should approximate the radial distribution function (RDF) of vortex lattice [53, 54],

$$f(r) = \sum_{n=1}^{\infty} \frac{N_n}{\sigma\sqrt{2\pi R_n/a_\Delta}} \exp\left[-\frac{(r-R_n)^2}{2\sigma^2 R_n/a_\Delta}\right], \qquad (2)$$

where $a_\Delta = \sqrt{2\Phi_0/\sqrt{3}H}$ is the expected lattice constant for a perfect hexagonal vortex lattice at the magnetic field of $H$, $\sigma \ll a_\Delta$ is the standard deviation of the discrepancy between nearest-neighbor distances, $R_n$ is the radius of the $n^{th}$ coordination shell, and $N_n$ is the number of sites in this shell. The oscillatory blue curves shows the fits of the experimental histograms to

$$Nf(r)\delta r \frac{\pi L^2 + (4-r)r^2 - 4rL}{\pi L^2}, \qquad (3)$$

with $N$ and $\delta r$ representing the involved vortex number and bin size of the histograms, respectively. The fractional term accounts for the finite image size. Here only two free parameters $R_1$ and $\sigma$ are relevant. Their values as well as the correlation length $\zeta$ of vortex lattices, estimated from $a_\Delta$ and $\sigma$ [53], are summarized in Table S2 [44]. The values of $\zeta$ are two or three times as large as $a_\Delta$ for varying film thickness and $H$, indicative of mediate pinning forces by the structural disorders in BaBi$_3$ films. It is worth noting that the fitted parameter $R_1$ deviates from $a_\Delta$ ($R_1 < a_\Delta$) with reduced film thickness and increasing $H$ (the bottom and middle panels in Fig. 4(e)). It might be because the fourfold Moiré ripples tend to configure vortices into square lattices, but are not strong enough to result in regular ones. This is evident by analysis of the autocorrelation images [Fig. S6]. Distinct from Fig. 4(c), the central rings are nearly square-shaped with the sides running along the directions of Moiré ripples.

**C. Correlation between disorder and superconducting gap**

In order to assess the impact of disorder on the superconducting gap, the tunneling conductance spectra on thin BaBi$_3$ films with varying thickness are taken and plotted in Fig. 5(a). As the film thickness is reduced, the increased structural disorders gradually suppress the coherence peaks, fill and sharpen the subgap DOS near $E_F$, leads eventually to V-shaped "soft" gap structure. These spectra look analogous to those widely observed in highly disordered superconductors [32, 33, 55-60], and deviate substantially from Eq. (1) with either isotropic or anisotropic s-wave gap functions [Fig. S7 and Table S3], as explained in the Supplementary Materials [44]. Recently, a model of finite-range superconducting fluctuations has been put forward for disordered superconductors [33, 61, 62]. In this model, instead of the effective energy broadening $\Gamma$ ($\Gamma = 0$) in Eq. (1), a superconducting correlation length $1/q_0$ is considered to calculate the tunneling spectrum. Interestingly, we find that all the experimental $dI/dV$ spectra in thin BaBi$_3$ films can be well described by the quasi-long-range superconducting fluctuation [61], namely

$$\Delta^2(V) = \frac{\Delta_0^2}{\pi^2} \frac{V}{\sqrt{v_F^2 q_0^2 + V^2}} \left[ \ln\left(\frac{\sqrt{v_F^2 q_0^2 + V^2} + V}{\sqrt{v_F^2 q_0^2 + V^2} - V}\right) - i\pi \right]. \qquad (4)$$

Here a small $v_F q_0$ means a slow spatial decay of pair function $\Delta(r)$. As the pair function $\Delta$ does not decay ($q_0 = 0$), one recovers the well-known BCS result. Figure 5(b) summarizes the film thickness-dependent $v_F q_0$ extracted from the best fits to Eq. 4 (blue curves in Fig. 5(a)), and disorder strength characterized by the root-mean-square (rms) deviation of STM image corrugation. Evidently, they reduce and scale with the film thickness in a power-law manner, e.g. $v_F q_0 \propto d^{-1.38}$. This establishes a direct linkage between the Moiré ripple-induced disorders and the gap softening. The disorders tend to increase the spatial decay of $\Delta(r)$ and suppress the coherence peaks in superconductors.

Notably, although the fluctuating model quantitatively follows the experimental d$I$/d$V$ spectra of BaBi$_3$ films with strong disorders, it lacks the validity by degrees with attenuating disorder on thick films, e.g. 11 UC film in Fig. 5(a), where surface stripes and disorders are intertwined and cooperatively affect the tunneling conductance spectra. The observation of slightly elongated vortex cores on such films supports this claim [Fig. 3(b)]. On the other hand, the Moiré ripples, despite being invisible from STM topography [Fig. 1(c)], yet play an essential role in superconductivity of the thickest films we investigated, justified by two observations. First, there remains a tiny discrepancy (e.g. near the gap edges) between the experimental d$I$/d$V$ curve and theoretical one in Fig. 2(b). From the fitting parameters of $\Delta_1 = 1.01$ meV and $\Delta_2 = 0.28$ meV, we extract a gap anisotropy of 1.77, larger than the anisotropy $\xi_b/\xi_a = 1.4$. This is primarily derived as a consequence of disorder, which kills the coherence peaks and leads to an overestimation of gap anisotropy in the theoretical fit. Second, an oblique vortex lattice is often observed for elongated vortices as the pinning is not effective [28, 46-48], due to a trade-off between the direction-dependent vortex-vortex interactions expected for elliptic vortices and the closet packing of vortices. However, imposed by the Moiré ripple-induced fourfold vortex pinning perturbation, this trade-off would be disturbed, leading to a nearly random arrangement of vortices, as revealed in Fig. 3(a). Further theoretical studies of anisotropic superconductors subject to strong disorder scattering may help fully understand the above tunneling spectra.

### D. Superconductivity and vortex core of SrBi$_3$

To gain further understanding regarding the interplay between symmetry breaking and superconductivity, we examine the tunneling d$I$/d$V$ spectrum at 0.4 K for another superconducting sister compound SrBi$_3$ with an almost identical $T_c$ (~ 5.6 K) for comparison. The substitution of Ba by Sr changes the crystal from tetragonal to cubic structure (space group P$m$-3$m$) with a lattice constant of ~ 5.055 Å [37, 38]. Figure 6(a) depicts an atomically-resolved STM image of 28 UC SrBi$_3$ (~ 14.2 nm) epitaxial films on SrTiO$_3$(001) substrate, exhibiting again a $\sqrt{2} \times \sqrt{2}$ surface reconstruction marked by the black square. Despite this similarity, one remarkable difference is that unlike BaBi$_3$ no surface stripe is observed in SrBi$_3$. Tunneling spectrum on SrBi$_3$, reveals a fully-gapped superconducting DOS that can be nicely fitted to the Dynes formula with an isotropic gap function $\Delta$ = 1.01 meV [Fig. 6(b)]. Moreover, circular-shaped vortex core [Fig. 6(c)] and clear ZBC peak [Fig. 6(d)] at the vortex core are demonstrated on SrBi$_3$ films when a magnetic field of 0.08 T is applied. This is in line with little disorder and negligible gap anisotropy involved in SrBi$_3$ films, which does not suffer from surface symmetry breaking.

**E. Atomic and electronic structure calculations of Ba(Sr)Bi$_3$**

We theoretically studied the atomic and electronic structures of Ba(Sr)Bi$_3$ surfaces by density functional theory. Let us first discuss about BaBi$_3$. The bulk shows a metallic band structure [Figs. 7(a) and 7(b)], as reported previously [37]. While 6$p$ orbitals of Ba are mostly located far above $E_F$ (i.e. unoccupied), 6$p$ orbitals of Bi are partially occupied and have a dominant contribution to states near $E_F$ [Fig. 7(a)]. Therefore transport-related properties are majorly determined by Bi, suggesting that the bright spots in STM topographies [Figs. 1(c) and 1(d)] can be assigned to Bi atoms.

The BaBi$_3$ bulk is composed of alternating layers of mixed Bi-Ba and pure Bi planes stacked along the (001) direction, giving two types of bulk-terminated (001) surfaces. Given that Ba (0.34 J/m$^2$) has a smaller surface energy than Bi (0.43 J/m$^2$) [63] and tends to enrich the surface, the observed termination should be the Bi-Ba mixed plane. Therefore, all possible $\sqrt{2} \times \sqrt{2}$ reconstructions of (001) surface can be constructed by introducing Bi vacancies or adatoms on the chemically mixed surface [Fig. S8]. These surface formation energy was calculated as a function of the chemical potential of Bi ($\mu_{Bi}$) by considering varying growth conditions. The two surface phases with 1/4 monolayer (ML) Bi vacancies or adatoms are most relevant to our experiments, since they give one Bi (i.e. bright spot) per surface supercell at the top layer as observed by STM. As shown in Fig. 7(c), the 1/4 ML vacancy (adatom) phase is prone to form at the Bi-poor (Bi-rich) condition. We cannot determine their relative stability without knowing the experimental chemical potential.

However, our calculations reveal that a homogeneous distribution of top-layer Bi is energetically more favorable than inhomogeneous ones only for the vacancy phase (Table S4), which enables the formation of a regular surface reconstruction. Moreover, the vacancy phase [Fig. 7(d)] gives a simulated STM image [Fig. 7(e)] consistent with experiment. These results thus imply that the Bi vacancy phase very likely corresponds to the surface observed experimentally.

Furthermore, the existence of Bi vacancies on the surface might explain the formation of surface stripe phases. Intuitively, the surface with Bi vacancies would experience a contractive strain that can deform the surface lattice. To determine the possible lattice deformation, we calculated the surface elastic constants of the Ba-Bi terminated surface along the high-symmetric $a$ (or $b$) and $a_0$ (or $b_0$) directions (defined in Fig. 1) and obtained $C_{aa} = C_{bb} = 34.8 \text{ GPa} \cdot \text{nm}$ and $C_{a_0 a_0} = C_{b_0 b_0} = 56.4 \text{ GPa} \cdot \text{nm}$. This result suggests that the surface lattice deformation is not isotropic but more easily happens along either the $a$ or $b$ direction, which lowers the surface rotational symmetry from $C_4$ to $C_2$, in agreement with our experiments. Therefore, the surface contractive strain induced by Bi vacancies is possibly a driving force of creating surface stripes. One might wonder whether the step edges of $SrTiO_3$ would impose additional strain and distort the surface stripes of $BaBi_3$. Our observations show that this factor is quite small as compared to the Bi vacancies, and never affect the surface stripes across the step edges.

The sister compound $SrBi_3$ shares similar atomic and electronic structures with $BaBi_3$ in the bulk, but displays significantly different features on the surface [Fig. S9], which can be understood by the different binding affinity of Bi on the $SrBi_3$ and $BaBi_3$ surfaces. The calculated adsorption energy of a Bi adatom on the Sr-Bi terminated surface is ∼ 0.18 eV higher than on the Ba-Bi surface. The stronger binding of Bi on $SrBi_3$ leads to two important consequences: i) The surface phase with 1/4 ML Bi adatoms is more stable than the one with 1/4 ML Bi vacancies over a broader range of $\mu_{Bi}$ than for $BaBi_3$ [cf. Fig. 7(c) and Fig. S9(b)], making the adatom phase [Fig. S9(c)] more likely appear in experiment; ii) Unlike $BaBi_3$, the adatom phase favors a homogeneous distribution, while the vacancy phase does not (Table S4). The adatom phase shows STM images in agreement with our experiment [Fig. S9(d)]. Therefore, we surmise that the type of surface defects varies from Bi vacancy to Bi adatom as the coupling between Bi and substrate gets stronger from $BaBi_3$ to $SrBi_3$. The distinct STM images of point defects observed on $BaBi_3$ [Fig. 1(d)] and $SrBi_3$ [Fig. 6(a)] surfaces seem to support this scenario. Furthermore, the adatom phase would experience no surface contractive strain, which might be the reason for the absence of surface stripe on $SrBi_3$.

## IV. CONCLUSIONS

Our thorough tunneling spectra, vortex core imaging and their thickness dependence, together with the comparison experiments with sister superconductor $SrBi_3$ compellingly demonstrate that superconductivity on symmetry-breaking $BaBi_3$ surface is in consistency with BCS predications for twofold symmetric order parameter, and that the superconducting pairing gaps in thin films with increased disorders appear more like V-shaped and are fairly accounted for by a model of fluctuating superconductivity. Our results show that the superconducting DOS of a fully opened *s*-wave gap could be profoundly modified by the symmetry breaking and disorders. This must be taken seriously for the interpretation and modelling of STM tunneling conductance spectra in superconductors with the coexistence of symmetry breaking and disorders, including high-$T_c$ cuprates and iron-based compounds, because their possible sign-changing superconductivity is more fragile against disorders and symmetry breaking states. This study thus may contribute to understand the tunneling conductance spectra in superconductors with rich physics of symmetry breaking states and disorders.

## Acknowledgements

This work is financially supported by the Ministry of Science and Technology of China (Grants No. 2017YFA0304600, 2016YFA0301004, 2016YFA0301001) and the National Natural Science Foundation of China (Grants No. 11774192, 11427903, 11504196, 11634007, 11674188, 11334006). C. L. S. and Y. X. acknowledge support from the National Thousand-Young-Talents Program and Tsinghua University Initiative Scientific Research Program.

W. L. Wang, Y. M. Zhang, and N. N. Luo contributed equally to this work.

**References**

[1] D. E. Moncton, J. D. Axe, and F. J. DiSalvo, Phys. Rev. B **16**, 801 (1977).

[2] A. M. Gabovich, A. I. Voitenko, J. F. Annett, and M. Ausloos, Supercond. Sci. Tech. **14**, R1 (2001).

[3] J. E. Hoffman, E. W. Hudson, K. M. Lang, V. Madhavan, H. Eisaki, S. Uchida, and J. C. Davis, Science **295**, 466 (2002).

[4] C. Pfleiderer, Rev. Mod. Phys. **81**, 1551 (2009).

[5] R. Comin, A. Frano, M. M. Yee, Y. Yoshida, H. Eisaki, E. Schierle, E. Weschke, R. Sutarto, F. He, A. Soumyanarayanan, Y. He, M. Le Tacon, I. S. Elfimov, J. E. Hoffman, G. A. Sawatzky, B. Keimer, A. Damascelli, Science **343**, 390 (2014).

[6] S. Margadonna, Y. Takabayashi, M. T. McDonald, M. Brunelli, G. Wu, R. H. Liu, X. H. Chen, and K. Prassides, Phys. Rev. B **79**, 014503 (2009).

[7] M. A. Beno, L. Soderholm, D. W. Capone, D. G. Hinks, J. D. Jorgensen, J. D. Grace, I. K. Schuller, C. U. Segre, and K. Zhang, Appl. Phys. Lett. **51**, 57 (1987).

[8] Y. Ando, K. Segawa, S. Komiya, and A. N. Lavrov, Phys. Rev. Lett. **88**, 137005 (2002).

[9] M. J. Lawler, K. Fujita, J. Lee, A. R. Schmidt, Y. Kohsaka, C. K. Kim, H. Eisaki, S. Uchida, J. C. Davis, J. P. Sethna and E. A. Kim, Nature **466**, 347 (2010).

[10] T. M. Chuang, M. P. Allan, J. Lee, Y. Xie, N. Ni, S. L. Bud'ko, G. S. Boebinger, P. C. Canfield, and J. C. Davis. Science **327**, 181 (2010).

[11] J. H. Chu, J. G. Analytis, K. De Greve, P. L. McMahon, Z. Islam, Y. Yamamoto, and I. R. Fisher, Science **329**, 824 (2010).

[12] M. Yi, D. H. Lu, J. H, Chu, J. G. Analytis, A. P. Sorini, A. F. Kemper, B. Moritz, S. K. Mo, R. G. Moore, M. Hashimoto, W. S. Lee, Z. Hussain, T. P. Devereaux, I. R. Fisher, Z. X. Shen, Proc. Natl Acad. Sci. U.S.A. **108**, 6878 (2011).

[13] S. Kasahara, H. J. Shi, K. Hashimoto, S. Tonegawa, Y. Mizukami, T. Shibauchi, K. Sugimoto, T. Fukuda, T. Terashima, A. H. Nevidomskyy and Y. Matsuda, Nature **486**, 382 (2012).

[14] K. Nakayama, Y. Miyata, G. N. Phan, T. Sato, Y. Tanabe, T. Urata, K. Tanigaki, and T. Takahashi, Phys. Rev. Lett. **113**, 237001 (2014).

[15] S. A. Kivelson, I. P. Bindloss, E. Fradkin, V. Oganesyan, J. M. Tranquada, A. Kapitulnik, and C. Howald, Rev. Mod. Phys. **75**, 1201 (2003).

[16] C. V. Parker, P. Aynajian, E. H. da Silva Neto, A. Pushp, S. Ono, J. S. Wen, Z. J Xu, G. D. Gu, and A. Yazdani, Nature **458**, 677 *(2010).*

[17] A. Soumyanarayanan, M. M. Yee, Y. He, J. van Wezel, D. J. Rahn, K. Rossnagele, E. W. Hudson, M. R. Norman, and J. E. Hoffman, Proc. Natl Acad. Sci. U.S.A. **110**, 1623 (2013).

[18] P. Giraldo-Gallo, Y. Zhang, C. Parra, H. C. Manoharan, M. R. Beasley, T.H. Geballe, M. J. Kramer, and I. R. Fisher, Nat. Commun. **6**, 8231 (2015).

[19] E. Morosan, H. W. Zandbergen, B. S. Dennis, J. W. G. Bos, Y. Onose, T. Klimczuk, A. P. Ramirez, N. P. Ong, and R. J. Cava, Nat. Phys. **2**, 544 (2006).

[20] B. Keimer, S. A. Kivelson, M. R. Norman, S. Uchida, and J. Zaanen, Nature **518**, 179 (2015).

[21] R. Comin, R. Sutarto, E. H. da Silva Neto, L. Chauviere, R. Liang, W. N. Hardy, D. A. Bonn, F. He, G. A. Sawatzky, and A. Damascelli, *Science* **347**, 1335 (2015).

[22] J. Wu, A. T. Bollinger, X. He, and I. Božović, Nature **547**, 432 (2017).

[23] F. Ronning, T. Helm, K. R. Shirer, M. D. Bachmann, L. Balicas, M. K. Chan, B. J. Ramshaw, R. D. McDonald, F. F. Balakirev, M. Jaime, E. D. Bauer, P. J. W. Moll, Nature **548**, 313 (2017).

[24] B. X. Zheng, C. M. Chung, P. Corboz, G. Ehlers, M. P. Qin, R. M. Noack, H. Shi, S. R. White, S. W. Zhang, and G. K. Chan, Science **358**, 1155 (2017).

[25] E. W. Huang, C. B. Mendl, S. X. Liu, S. Johnston, H. C. Jiang, B. Moritz, and T. P. Devereaux, Science **358**, 1161 (2017).

[26] J. Chang, E. Blackburn, A. T. Holmes, N. B. Christensen, J. Larsen, J. Mesot, R. X. Liang, D. A. Bonn, W. N. Hardy, A. Watenphul, M. V. Zimmermann, E. M. Forgan, S. M. Hayden, Nat. Phys. **8**, 871 (2012).

[27] I. Guillamón, H. Suderow, S. Vieira, L. Cario, P. Diener, and P. Rodière, Phys. Rev. Lett. **101**, 166407 (2008).

[28] C. L. Song, Y. L.Wang, P. Cheng, Y. P. Jiang, W. Li, T. Zhang, Z. Li, K. He, L. L. Wang, J. F. Jia, H. H. Hung, C. J. Wu, X. C. Ma, X. Chen, Q. K. Xue, Science **332**, 1410 (2011).

[29] R. M. Fernandes and A. J. Millis, Phys. Rev. Lett. **111**, 127001 (2013).

[30] J. A. Slezak, J. Lee, M. Wang, K. McElroy, K. Fujita, B. M. Andersen, P. J. Hirschfeld, H. Eisaki, S. Uchida, and J. C. Davis, Proc. Natl Acad. Sci. U.S.A. **105**, 3203-3208 (2013).

[31] I. Zeljkovic, Z. J. Xu, J. S. Wen, G. D. Gu, R. S. Markiewicz, and J. E. Hoffman, Science **337**, 320 (2012).

[32] C. Brun, T. Cren, V. Cherkez, F. Debontridder, S. Pons, D. Fokin, M. C. Tringides, S. Bozhko, L. B. Loffe, B. L. Altshuler, D. Roditchev, Nat. Phys. **10**, 444 (2014).

[33] S. Takei, B. M. Fregoso, H. Y. Hui, A. M. Lobos, and S. Das Sarma, Phys. Rev. Lett. **110**, 186803 (2013).

[34] K. Luna, P. Giraldo-Gallo, T. Geballe, I. Fisher, and M. Beasley, Phys. Rev. Lett. **113**, 177004 (2014).

[35] Y. Zhong, Y. Wang, S. Han, Y. F. Lv, W. L. Wang, D. Zhang, H. Ding, Y. M. Zhang, L. L. Wang, K. He, R. D. Zhong, J. A. Schneeloch, G. D. Gu, C. L. Song, X. C. Ma, Q. K. Xue, Sci. Bull. **61**, 1239 (2016).

[36] N. Haldolaarachchige, S. K. Kushwaha, Q. Gibson, and R. J. Cava, Supercond. Sci. Tech. **27**, 105001 (2014).

[37] D. F. Shao, X. Luo, W. J. Lu, L. Hu, X. D. Zhu, W. H. Song, X. B. Zhu, and Y. P. Sun, Sci. Rep. **6**, 21484 (2016).

[38] R. Jha, M. A. Avila, and R. A. Ribeiro, Supercond. Sci. Tech. **30**, 025015 (2017).

[39] G. Kresse and J. Furthmüller, Phys. Rev. B **54**, 11169 (1996).

[40] J. P. Perdew, K. Burke, and M. Ernzerhof, Phys. Rev. Lett. **77**, 3865 (1996).

[41] G. Kresse and D. Joubert, Phys. Rev. B **59**, 1758 (1999).

[42] J. Tersoff and D. R. Hamann, Phys. Rev. B **31**, 805 (1985).

[43] Y. F. Lv, W. L. Wang, Y. M. Zhang, H. Ding, W. Li, L. L. Wang, K. He, C. L. Song, X. C. Ma, and Q. K. Xue, Sci. Bull. **62**, 852 (2017).

[44] See Supplemental Materials for details on Moiré superstructure, surface stripes, theoretical fits and magnetic vortices.

[45] R. C. Dynes, V. Narayanamurti, and J. P. Garno, Phys. Rev. Lett. **41**, 1509 (1978).

[46] I. Maggio-Aprile, Ch. Renner, A. Erb, E. Walker, and Ø. Fischer, Phys. Rev. Lett. **75**, 2754 (1995).

[47] Q. Fan, W. H. Zhang, X. Liu, Y. J. Yan, M. Q. Ren, M. Xia, H. Y. Chen, D. F. Xu, Z. R. Ye, W. H. Jiao, G. H. Cao, B. P. Xie, T. Zhang and D. L. Feng, Phys. Rev. B **91**, 104506 (2015).

[48] Z. Y. Du, D. Fang, Z. Y. Wang, Y. F. Li, G. Du, H. Yang, X. Y. Zhu, and H. H. Wen, Sci. Rep.

5, 9408 (2015).

[49] W. L. Wang, Y. M. Zhang, Y. F. Lv, H. Ding, L. L. Wang, W. Li, K. He, C. L. Song, X. C. Ma, Q. K. Xue, Phys. Rev. B 97, 134524 (2018).

[50] C. Caroli, P. G. De Gennes, and J. Matricon, Phys. Lett. **9**, 307 (1964).

[51] Ch. Renner, A. D. Kent, Ph. Niedermann, Ø. Fischer, and F. Lévy, Phys. Rev. Lett. **67**, 1650 (1991).

[52] Y. X. Ning, C. L. Song, Y. L. Wang, X. Chen, J. F. Jia, Q. K. Xue, and X. C. Ma, J. Phys.: Condens. Matter **22**, 065701 (2010).

[53] D. S. Inosov, T. Shapoval, V. Neu, U. Wolff, J. S. White, S. Haindl, J. T. Park, D. L. Sun, C. T. Lin, E. M. Forgan, M. S. Viazovska, J. H. Kim, M. Laver, K. Nenkov, O. Khvostikova, S. Kühnemann, and V. Hinkov, Phys. Rev. B **81**, 014513 (2010).

[54] C. L. Song, Y. Yin, M. Zech, T. Williams, M. M. Yee, G. F. Chen, J. L. Luo, N. L. Wang, E. W. Hudson, and J. E. Hoffman, Phys. Rev. B 87, 214519 (2013).

[55] S. P. Chockalingam, M. Chand, Anand Kamlapure, J. Jesudasan, A. Mishra, V. Tripathi, and P. Raychaudhuri, Phys. Rev. B **79**, 094509 (2009).

[56] B. Sacépé, C. Chapelier, T. I. Baturina, V. M. Vinokur, M. R. Baklanov, and M. Sanquer, Nat. Commun. **1**, 140 (2010).

[57] M. Mondal, A, Kamlapure, M. Chand, G. Saraswat, S. Kumar, J. Jesudasan, L. Benfatto, V. Tripathi, and P. Raychaudhuri, Phys. Rev. Lett. **106**, 047001 (2011).

[58] B. Sacépé, T. Dubouchet, C. Chapelier, M. Sanquer, M. Ovadia, D. Shahar, M. Feigel'man, and L. Ioffe, Nat. Phys. **7**, 239 (2011).

[59] C. Richter, H. Boschker, W. Dietsche, E. Fillis-Tsirakis, R. Jany, F. Loder, L. F. Kourkoutis, D. A. Muller, J. R. Kirtley, C. W. Schneider and J. Mannhart, Nature **502**, 528 (2013).

[60] Y. Noat, V. Cherkez, C. Brun, T. Cren, C. Carbillet, F. Debontridder, K. Ilin, M. Siegel, A. Semenov, H.-W. Hübers, and D. Roditchev, Phys. Rev. B **88**, 014503 (2013).

[61] D. Dentelski, A. Frydman, E. Shimshoni, and E. G. Dalla Torre, Phys. Rev. B **97**, 100503 (2018).

[62] I. S. Burmistrov, I. V. Gornyi, and A. D. Mirlin, Phys. Rev. B **93**, 205432 (2016).

[63] B. J. Keene, Int. Mater. Rev. **38**, 157 (2013).

**Figure captions**

**Fig. 1.** Morphology and surface structure of BaBi$_3$ films. (a) Schematic crystal structure of non-oxide perovskite superconductor BaBi$_3$. The *a*, *b* and *c* axes are aligned along the crystallographic orientations. (b) STM topography ($V$ = 1.5 V, $I$ = 30 pA, 300 nm × 300 nm) of ~ 22 UC thick BaBi$_3$ epitaxial films on SrTiO$_3$(001). The principle directions of SrTiO$_3$(001) surface run along the $a_0$ and $b_0$ axes. (c) A zoom-in view of BaBi$_3$(001) surface ($V$ = 5 mV, $I$ = 100 pA, 16 nm × 16 nm) exhibiting stripe-like order. The black square indicates the unit cell of $\sqrt{2} \times \sqrt{2}$ reconstructed surface. Every bright spots denote the top Bi atoms. (d) Atomic resolution STM image of BaBi$_3$ films grown under Bi-richer atmosphere ($V$ = -30 mV, $I$ = 130 pA, 16 nm × 16 nm), presenting Bi-site adatoms (see the overlaid Bi balls). The surface strips are visually overcome by addition of high-contrast defects. (e) STM image on 5 UC films ($V$ = 10 mV, $I$ = 200 pA, 100 nm × 100 nm), exhibiting the coexistence of surface stripe and Moiré superstructure.

**Fig. 2.** Tunneling spectroscopy. (a) Spatially-resolved differential conductance d$I$/d$V$ spectra at 0.4 K, taken on a 272 nm × 272 nm (4 × 4 grid) region of ~ 34 UC BaBi$_3$ films, illustrating the spatial homogeneity. (b) Normalization of the averaged d$I$/d$V$ spectra (empty circles) and its best fits to the Dynes formula with *s*-wave (magenta line) and anisotropic *s*-wave (blue line) gap functions, respectively. The experimental data shows a better fit to the twofold symmetric gap function $\Delta$ = 1.01 + 0.28cos(2θ) meV. (c) Equally spaced d$I$/d$V$ spectra, and (d) normalized d$I$/d$V$ spectrum as well as its best fit to a single *s*-wave gap on 5UC films.

**Fig. 3.** Vortex core structure and bound states. (a-c) Thickness-dependent internal structure of vortex cores on BaBi$_3$ films, with the thickness $d$ of (a) 34 UC, (b) 11 UC and (c) 5 UC. The applied magnetic field and STM image sizes are (a, b) 0.2 T, 272 nm × 272 nm and (c) 0.5 T, 182 nm × 182 nm, respectively. Dashed ellipses or circles draw the periphery of vortex cores, showing a transition from elliptical to circular vortex core with reduced film thickness. The white dashed lines designate grain boundaries, separating regions with orthogonal orientations of surface stripes and vortices. (d) High-resolution STM topography ($V$ = 10 mV, $I$ = 200 pA, 73 nm × 73 nm) on 34 UC films, and (e) simultaneous ZBC map showing a single vortex at 0.05 T, revealing the vortex elongation along the surface stripes. The thin white lines mark the surface stripes for easy viewing. (f) Radial dependence of ZBC in (e), and the fits to the Ginzburg-Landau theory. (g) Tunneling d$I$/d$V$ spectra measured at equally spaced (2.8 nm) distance across the vortex core of (e), exhibiting clear bound states at the vortex center of 34 UC films (magenta curve). (h) d$I$/d$V$ spectra across a vortex core of 5 UC films, presenting no ZBC peak. Setpoint: $V = 10$ mV, $I = 100$ pA.

**Fig. 4.** Vortex configuration. (a, b) Magnetic vortex lattices (454 nm × 454 nm) of ~ 20 UC thick BaBi$_3$ films at 0.2 T and 0.4 T, respectively. Voronoi cells (white lines) are overlaid onto the vortex maps. The white dashes mark one high-resolution vortex core that is also elongated along the surface stripes. (c) Two-dimensional autocorrelation function calculated from vortex map in (a), indicative of distorted hexagonal lattice (black polygon). (d) Delaunay triangulation diagram (gray lines) of the vortex map in (b). Every vortex has been color coded based on its coordination number (yellow, 4; cyan, 5; blue, 6; green, 7). (e) Histograms of vortex pair distances $d_{ij}$ at varying film thickness and field. Blue lines show the best fits of the experimental RDF to Eq. (3).

**Fig. 5.** Correlation between superconducting gap and disorder. (a) Film thickness dependence of d$I$/d$V$ spectra on BaBi$_3$ thin films. Blue lines show the best fits to the model of quasi-long-range superconducting fluctuations in Eq. (4). The spectra have been vertically offset for clarity, with the zero conductance positions marked by the correspondingly colored horizontal lines. (b) Double-logarithmic plots of fluctuation strength $v_F q_0$ (bottom panel) and rms deviation (top panel) caused by the Moiré ripple *versus* film thickness $d$. They scale as $d^{-0.38 \pm 0.08}$ and $d^{-0.54 \pm 0.03}$ with $d$. The error bars denote the standard deviation from 60 STM images and fits of 50 d$I$/d$V$ spectra.

**Fig. 6.** STM characterization of SrBi$_3$ films. (a) Atomically-resolved STM image ($V = 5$ mV, $I = 100$ pA, 16 nm × 16 nm) on ~ 28 UC SrBi$_3$ epitaxial films, presenting two Bi-site vacancies. The black square marks the $\sqrt{2} \times \sqrt{2}$ surface reconstruction, with a periodicity of about 7.2 Å. (b) Tunneling spectrum ($V = 5$ mV and $I = 100$ pA) of SrBi$_3$ (black circles), and the best fit to the Dynes formula in Eq. (1), with an isotropic *s*-wave pairing gap (magenta line). (c) ZBC map showing a circular-shaped vortex core (black dashes) on SrBi$_3$. (d) Bound state at the vortex center. Set point: $V = 5$ mV and $I = 100$ pA.

**Fig. 7.** DFT calculations of BaBi$_3$. (a) Electronic band structure and DOS of bulk BaBi$_3$. (b) The Brillouin zone of bulk BaBi$_3$ and high symmetry K points. (c) Surface formation energy ($\Delta E$) of $\sqrt{2} \times \sqrt{2}$ reconstructed BaBi$_3$(001) surfaces as a function of the chemical potential of Bi,

referenced to the Ba-Bi terminated surface, with $\mu_{Bi} = 0$ corresponding to bulk Bi. (d) Top view atomic structure of the Ba-Bi terminated surface with 1/4 ML Bi vacancies, and (e) its simulated STM image taken at 0.1 eV above $E_F$. A surface supercell is outlined by red dashed lines. Atoms in deep layers are shaded for clarity.

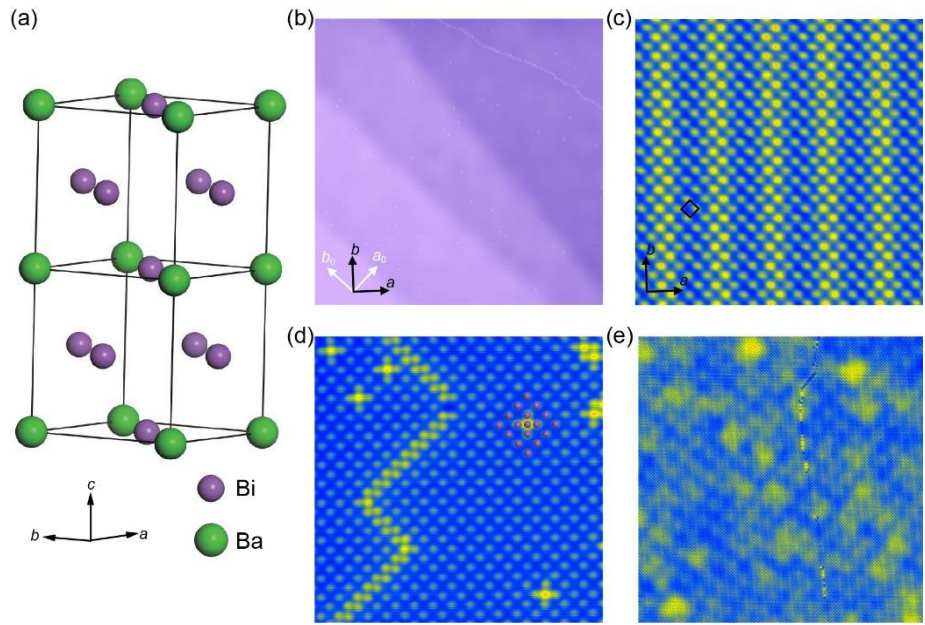

**Figure 1**

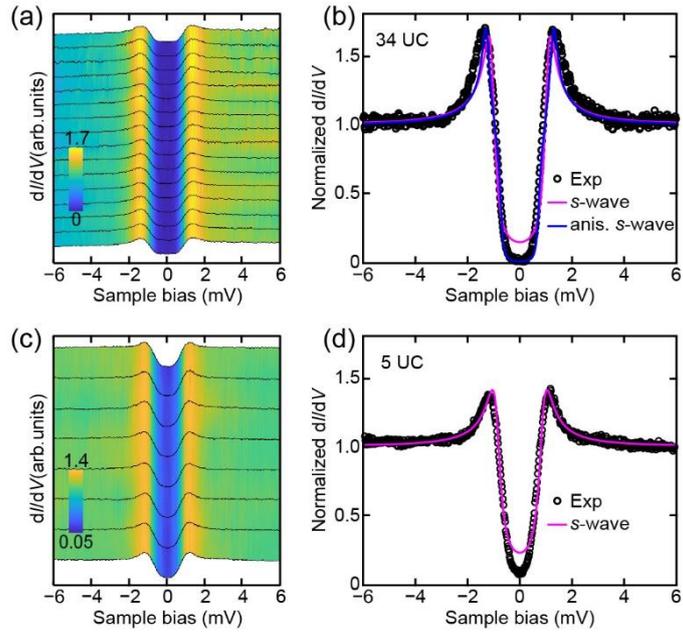

**Figure 2**

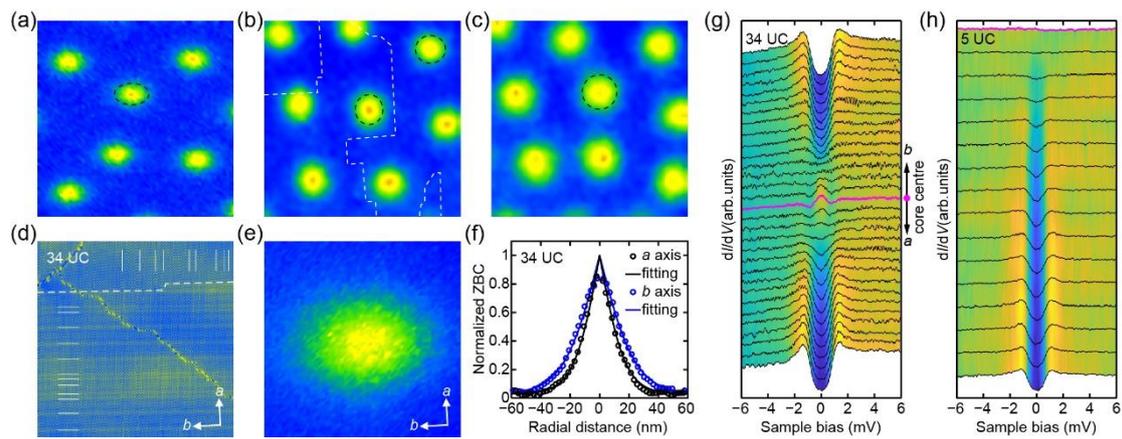

**Figure 3**

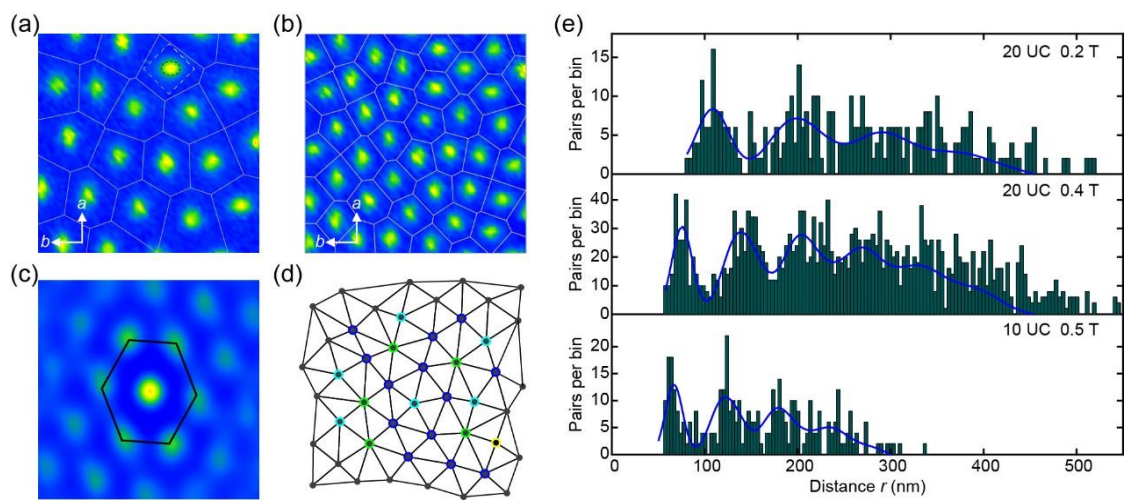

**Figure 4**

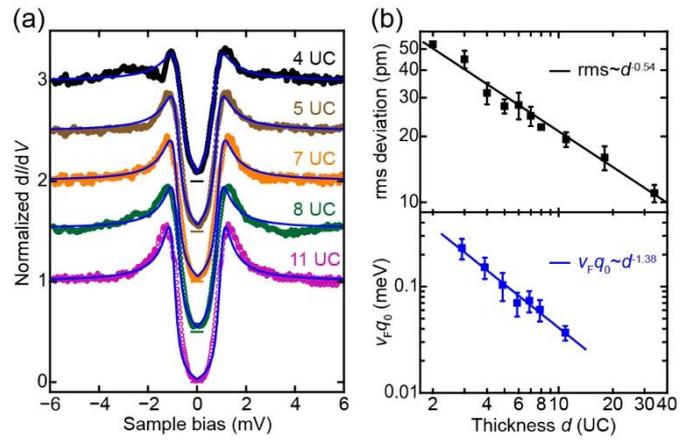

**Figure 5**

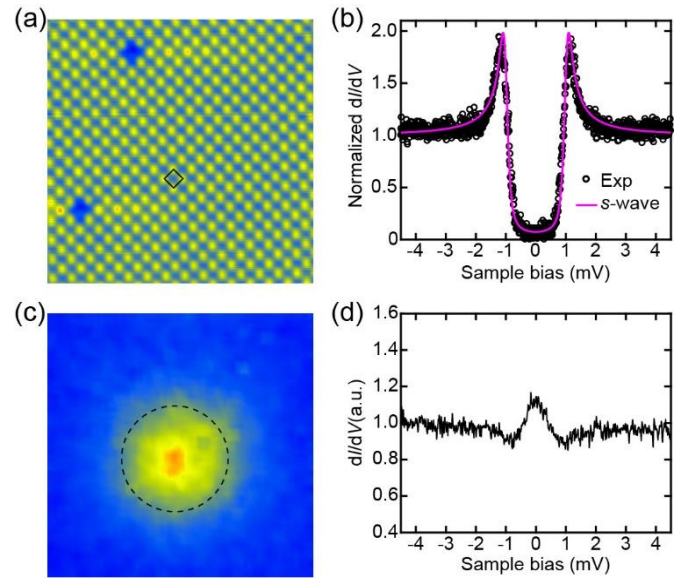

**Figure 6**

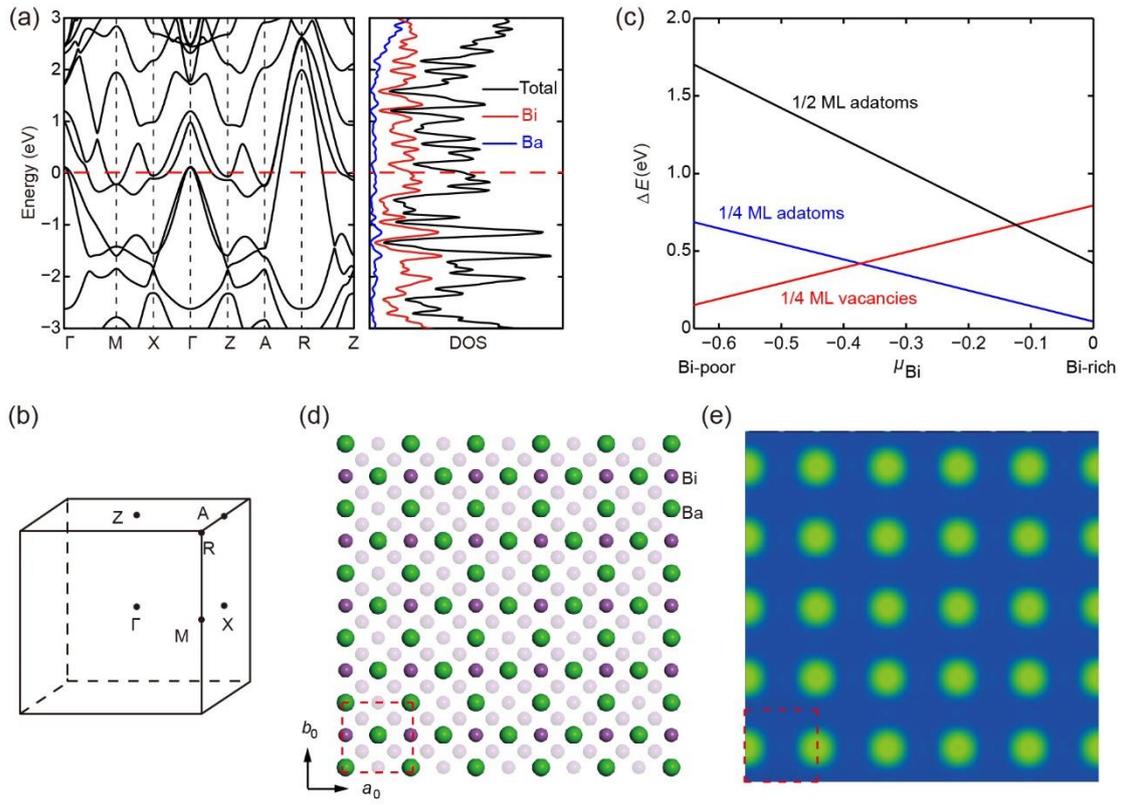

**Figure 7**

## Supplemental Material for:

**Surface symmetry breaking and disorder effects on superconductivity in perovskite BaBi$_3$ epitaxial films**

Wen-Lin Wang[1], Yi-Min Zhang[1], Nan-Nan Luo[1], Jia-Qi Fan[1], Chong Liu[1], Zi-Yuan Dou[1], Lili Wang[1,2], Wei Li[1,2], Ke He[1,2], Can-Li Song[1,2,*], Yong Xu[1,2,3,†], Wenhui Duan[1,2,4], Xu-Cun Ma[1,2], Qi-Kun Xue[1,2,‡]

[1]*State Key Laboratory of Low-Dimensional Quantum Physics, Department of Physics, Tsinghua University, Beijing 100084, China*

[2]*Collaborative Innovation Center of Quantum Matter, Beijing 100084, China*

[3]*RIKEN Center for Emergent Matter Science (CEMS), Wako, Saitama 351-0198, Japan*

[4]*Institute for Advanced Study, Tsinghua University, Beijing 100084, China*

**Formation of Moiré superstructure of epitaxial BaBi$_3$ films on SrTiO$_3$**

Shown in Figs. S1(a-c) are the varying sized STM topographies of BaBi$_3$ thin films ($\leq$ 7 UC) on SrTiO$_3$(001) substrate. In addition to the surface stripes [Figs. S1(b) and S3], square ripples, which we justify as Moiré superstructure between the epitaxial BaBi$_3$ overlayer and SrTiO$_3$(001) substrate, are apparent [Figs. S1(b) and S1(c)]. The Moiré superstructure-induced surface ripples are found to get faint with film thickness [Fig. 5(b)]. Based on the atomically-resolved STM image in Fig. S1(c), we notice that the Moiré lattice is rotated by $\sim 51°$ relative to the [0 1 0] azimuth of BaBi$_3$ crystal lattice. In order to extract the periodicity of Moiré superstructure, the two-dimensional fast Fourier transform (FFT) of STM image is performed and inserted in Fig. S1(d). Peaks which arise from the Moiré superstructure and $\sqrt{2} \times \sqrt{2}$ reconstruction are highlighted with green and white circles, respectively. By tracing the peak positions of FFT profiles [Fig. 1(d)], the periodicity $A$ of Moiré superstructure is estimated to be $\sim 66.5$ Å. Below we extend the so-called "hexagonal number sequence" method [1, 2] to square lattices, and explain how the Moiré fringe is formed based on the observed experimental parameters.

We mark the primitive vectors of underlying SrTiO$_3$(001) substrate as $a_0 = (1, 0)a_0$ and $b_0 = (0, 1)b_0$, with $a_0 = b_0 = 3.905$ Å denoting the lattice parameter of SrTiO$_3$(001). Any lattice vector of SrTiO$_3$(001) is thus a linear combination of the two primitive vectors, $\boldsymbol{R} = m_0\mathbf{a}_0 + n_0\mathbf{b}_0$, where $m_0$ and $n_0$ are integers. For commensurate superstructures on this substrate, if the primitive vectors of the BaBi$_3$ overlayer are represented by a = (1, 0)$a$ and b = (0, 1)$b$, they should satisfy $m_0\mathbf{a}_0 + n_0\mathbf{b}_0 = m\mathbf{a} + n\mathbf{b}$, where $m_0$, $n_0$, $m$ and $n$ are all integers. Each combination of $m_0$, $n_0$, $m$ and $n$ correspond to a commensurate Moiré superstructure with a certain periodicity and rotation angle $\theta$ between the two sets of primitive vectors. By listing all possible $m_0$, $n_0$, $m$ and $n$ combinations, one can easily identify the one best-matching with our experiments.

For simplicity, we first consider that the BaBi$_3$ thin films are terminated by the basal plane $c$(001), leading to $a = b \sim 5.188$ Å. In this case, the angles of Moiré superstructure with the SrTiO$_3$ substrate and BaBi$_3$ overlayer can be written as $\theta_{m,n} = \tan^{-1}\left(\frac{n}{m}\right)$ and $\theta_{m_0,n_0} = \tan^{-1}\left(\frac{n_0}{m_0}\right)$, respectively. The difference $\theta$ between them represents the intersection angle between the BaBi$_3$ overlayer and SrTiO$_3$ lattices, which is approximately 45° in experiment [Figs. 1(b) and S1(a-c)]. This imposes a strong constraint on the possible Moiré patterns. As summarized in Table S1 are the superstructures exhibiting the closer matches to both the experimental values $A \sim 66.5$ Å and $\theta \sim 45°$. In view of the experimental value of $\theta_{m,n} \sim 51°$, the Moiré superstructure best-matching our experiments is well judged and indicated by bold type in Table S1. Here the intersection angle $\theta = \theta_{m,n} - \theta_{m_0,n_0}$ is calculated to be 44.6°, in good consistency with the experimental observation. The Moiré superstructure can thus be denoted as SrTiO$_3$(001)-($\sqrt{293} \times \sqrt{293}$ )R6.7°. Furthermore, the ratio $R$ between the lattice parameters of SrTiO$_3$ and BaBi$_3$ can be calculated to be 0.74851 by $R = \sqrt{(m^2 + n^2)/(m_0^2 + n_0^2)}$, giving the lattice constant of 5.22 Å for epitaxial BaBi$_3$ films. This value is slightly larger than the theoretical one (5.188 Å), suggesting that the thin films might most probably be strained in tension. Drawn in Fig. S1(e) is the simulated Moiré superstructure, with its unit cell marked by the green square. We emphasize that the other equivalent Moiré superstructure, which relates to the original one by minor symmetry with respect to the primitive crystal axes (a$_0$ or b$_0$) of SrTiO$_3$, is in principle allowed and observed in Fig. 1(b). The two superstructures, with an intersection angle of 13.4°, coexist and destroy the long-range order of the Moiré ripples [Fig. 1(b)].

Now we explain how we can apply the Moiré superstructure to conclude the (001) termination of BaBi$_3$ films on SrTiO$_3$(001). As discussed in the main text of this manuscript, the BaBi$_3$ crystallizes into a tetragonal crystal structure. Previous studies reveal that in BaBi$_3$ the $c$-axis lattice constant is slightly smaller than the *a*(*b*)-axis one, with a difference δ ranging from 0.6%-1.4% [3-5]. As thus, the in-plane lattice constants along the two orthogonal axes will exhibit a small difference δ if the epitaxial BaBi$_3$ films is terminated by the basal plane *a*(001) or *b*(010). Figure S2 shows the simulated Moiré superstructures, which changes critically with δ. Although the square lattice (δ = 0) in the *a-b* plane leads to a square superstructure [Fig. 1(b)], the rectangular lattice with a small δ appreciably distorts the Moiré superstructure and gives rise to the rhombic patterns. A small discrepancy δ of 0.6% can distort the Moiré patterns by 7° [Fig. 2(a)], being readily distinguishable in experiment if it were present. In other words, the Moiré superstructure has a unique ability to magnify and discriminate any possible tiny lattice difference [6]. Thus the observed square Moiré ripples verify that the epitaxial BaBi$_3$ films on SrTiO$_3$(001) should be *c*(001)-terminated (δ = 0).

**Evaluating goodness of fit models to the experimental data**

In order to quantitatively evaluate the fit goodness of experimental data $Y_{\exp}$(V) by different models, we summarize the square of residuals between the observed values $Y_{\exp}$ and the values $Y_{\text{theory}}$ under the tested model, namely $\chi^2 = \Sigma_V(Y_{\exp} - Y_{\text{theory}})^2$. Enumerated in Table S3 are the resultant fitting parameters for best-following the experimental data under various models, i.e. the superconducting fluctuation model, Dynes model with isotropic and anisotropic *s*-wave gap functions. To better evaluate the goodness of fit, we have concentrated on the in-gap DOS (|V| < 1.1 mV) and calculated the discrepancies $\chi^2$ between theory and experiment in the energy range of ± 1.1 meV. As clearly demonstrated in Table S3, the fluctuating model gives the best fits of experimental d*I*/d*V* spectra on thin films, while the Dynes model with anisotropic *s*-wave gap functions explain the d*I*/d*V* spectra of thick BaBi$_3$ films. We emphasize that in thin films the values of $\chi^2$ are always smaller in the fluctuating model than those in anisotropic gap model, although the latter model involves more free parameters (Δ$_1$, Δ$_2$ and Γ) for the fits. Remarkably, $\chi^2$ appears negligibly small in the thin film limit, in which the Moiré ripples-induced disorder scattering get stronger. All these findings support the applicability of fluctuating model in disordered superconductors.

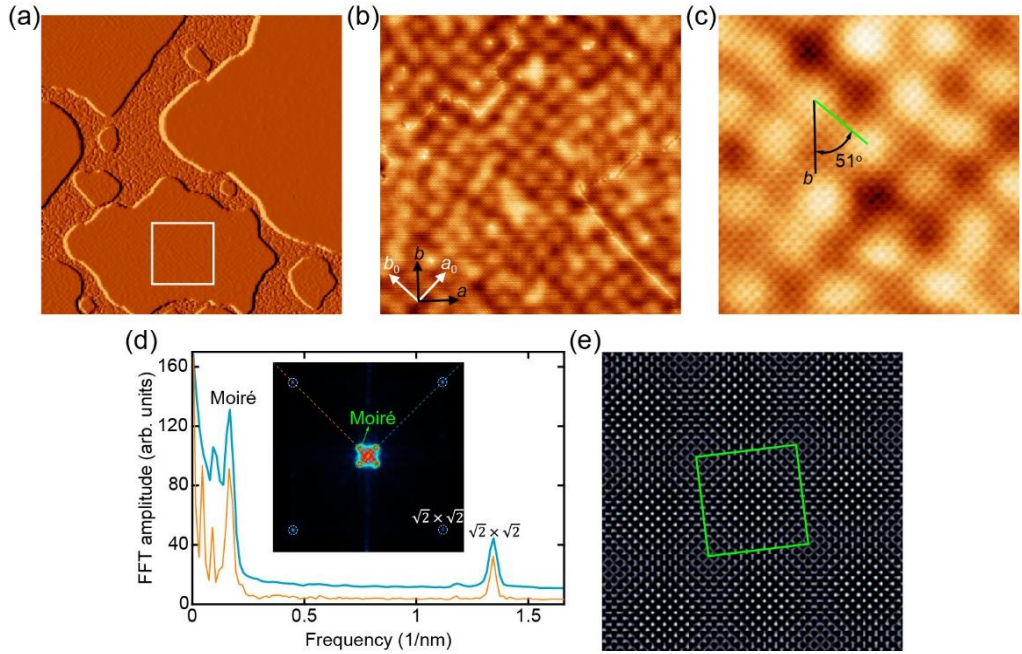

**Figure S1.** Moiré superstructure of $BaBi_3$ epitaxial films on $SrTiO_3(001)$. (a) Differential STM topographic image ($V = 1.5$ V, $I = 30$ pA, 454 nm × 454 nm) of $BaBi_3$ thin films on $SrTiO_3(001)$, revealing Morié-like fringes on the top surface. (b) Magnified STM image ($V = 10$ mV, $I = 30$ pA, 92 nm × 92 nm) on 5 UC film, acquired on a region marked by the white square in (a). (c) STM image ($V = 1.5$ V, $I = 200$ pA, 25 nm × 25 nm) showing the relative orientation between the Moiré superstructure (green line) and $BaBi_3$ crystal axis $b$ (black line). (d) Intensity profiles of fast Fourier transform (inset) of the STM image in (b), taken along the two orthogonal $a_0$ and $b_0$ axes. A moiré superstructure, with a periodicity of about 66.5 Å, is evident from both the profiles and FFT image (green circles). The square array of spots surrounded by the white circles correspond to the $\sqrt{2} \times \sqrt{2}$ surface reconstruction. (**e**) Simulated Moiré pattern based on the experimental parameters. The green square shows the unit cell of Moiré superstructure with a periodicity of 66.8 Å.

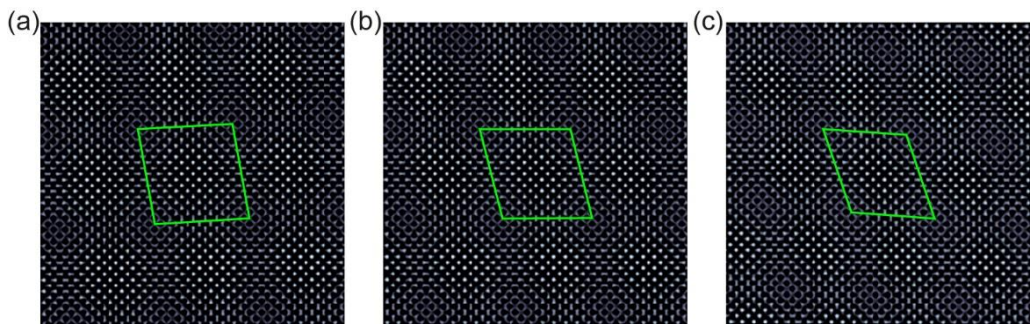

**Figure S2.** Moiré superstructure with varying in-plane lattice constant difference δ. The in-plane

lattice constants along the two orthogonal axes differ by (a) 0.6%, (b) 1.3% and (c) 2.5%. The green rhombuses mark the unit cells of Moiré superstructures.

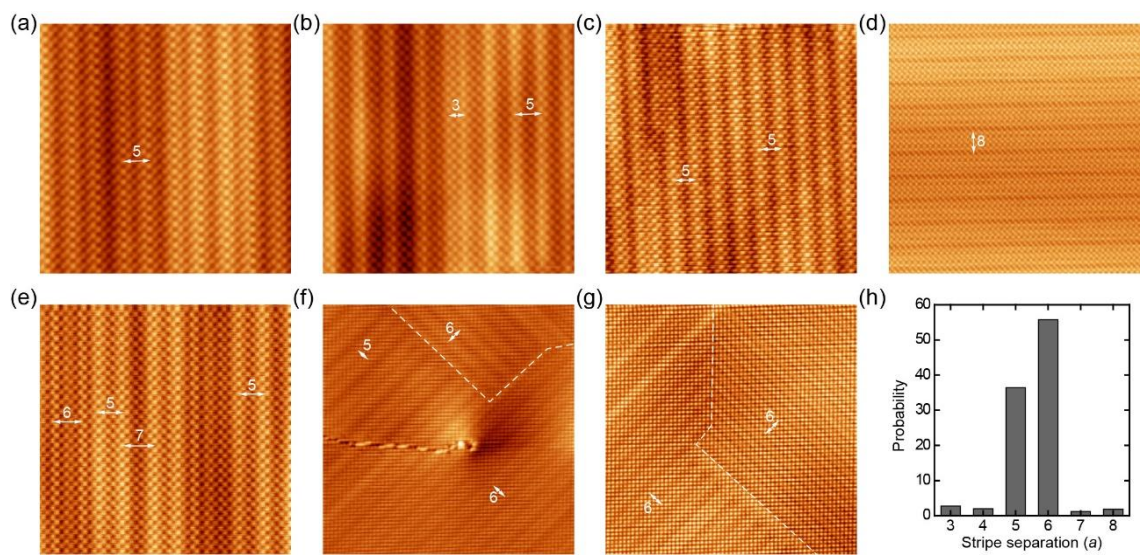

**Figure S3.** Devil's staircase-like surface stripe phases. (a-g) STM images exhibiting the universal existence of surface stripes on $BaBi_3$. The arrows and numbers mark the separation of neighboring stripes, in the unit of lattice constant $a$. The white lines designate grain boundaries, across which the stripe orientation alters by 90°. (h) Histograms showing the distribution of stripe width. The statistics involves roughly 130 STM images. Tunneling conditions and image size: (a) $V = 10$ mV, $I = 30$ pA, 28 nm × 28 nm; (b) $V = 5$ mV, $I = 100$ pA, 29 nm × 29 nm; (c) $V = 20$ mV, $I = 70$ pA, 29 nm × 29 nm; (d) $V = -10$ mV, $I = 30$ pA, 49 nm × 49 nm; (e) $V = 20$ mV, $I = 30$ pA, 28 nm × 28 nm; (f) $V = 10$ mV, $I = 20$ pA, 52 nm × 52 nm; and (g) $V = 10$ mV, $I = 200$ pA, 46 nm × 46 nm.

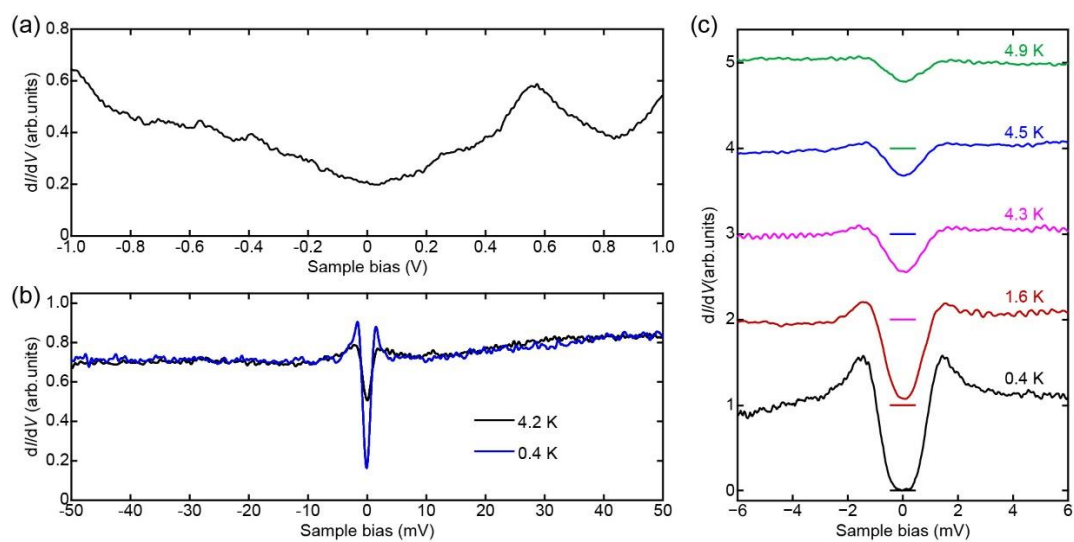

**Figure S4.** Tunneling d$I$/d$V$ spectra in various energy ranges. (a) Large-energy-scale d$I$/d$V$ spectrum on BaBi$_3$ films characteristic of finite DOS near $E_F$ and metallic behavior, regardless of film thickness. (b, c) Temperature-dependent d$I$/d$V$ spectra of 34 UC BaBi$_3$ films. For clarity the d$I$/d$V$ spectra in (c) have been vertically offset, with their zero conductance positions marked by correspondingly colored horizontal lines. At elevated temperatures, the subgap DOS is gradually lifted and the coherence peaks are suppressed, characteristic of superconductivity.

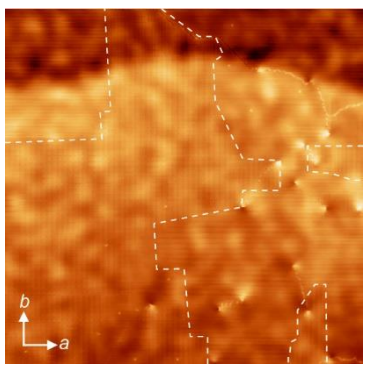

**Figure S5.** A 272 nm × 272 nm topographic image ($V$ = 10 mV, $I$ = 200 pA) where the vortex map of Fig. 3(b) is measured. The white dashed lines designate grain boundaries, separating regions with orthogonal orientations of surface stripes.

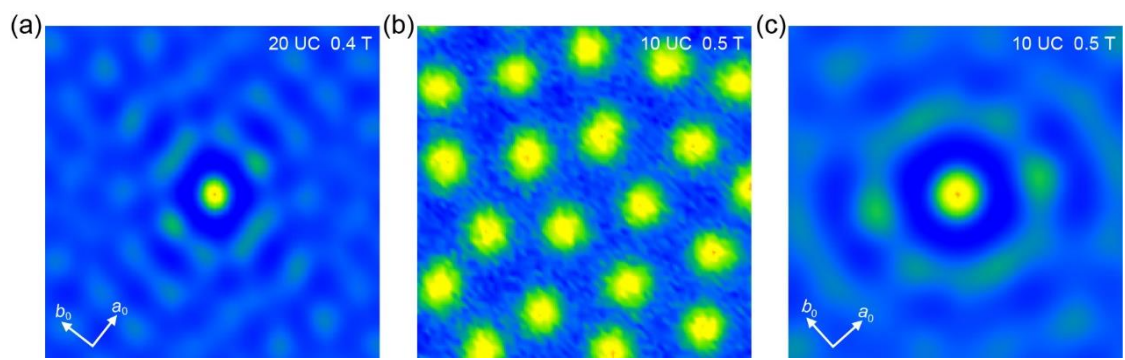

**Figure S6.** Vortex lattice and autocorrelation. (a) Two-dimensional autocorrelation function calculated from vortex lattice in Fig. 4b. (b) Vortex configuration of 10 UC films at a higher magnetic field of 0.5 T. The image size is 272 nm × 272 nm. (c) Autocorrelation function calculated from (b). The emergent four-fold symmetric features in (a) and (b) and local square vortex lattices in (b) reveal the roles played by the fourfold Moiré superstructure-induced ripples.

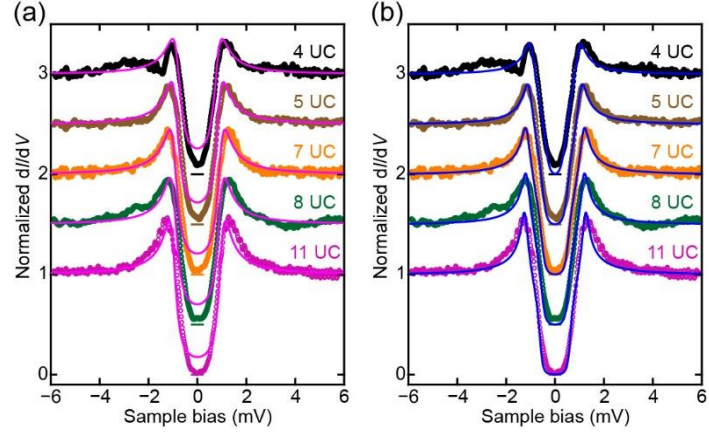

**Figure S7.** Fits of the experimental data by the Dynes model. (a) Film thickness-dependent d$I$/d$V$ spectra and their best fits to isotropic *s*-wave gap function (magenta curves). (b) The same series of d$I$/d$V$ spectra fitted by anisotropic *s*-wave gap function (blue curves).

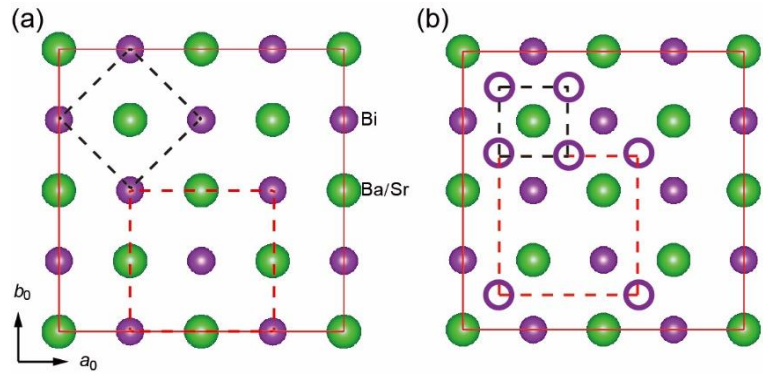

**Figure S8.** Theoretical study on possible distributions of surface defects. Inhomogeneous (black dashed squares) and homogeneous (red dashed squares) distribution of surface defects in a $2\sqrt{2} \times 2\sqrt{2}$ supercell for 1/4 ML Bi (a) vacancies and (b) adatoms, respectively. Surface Bi defects are created on the sites along the squares. Bi adatoms prefer to adsorb at the hollow site, denoted by open violet circles in (b). Only the top layer of the surface is shown for clarity.

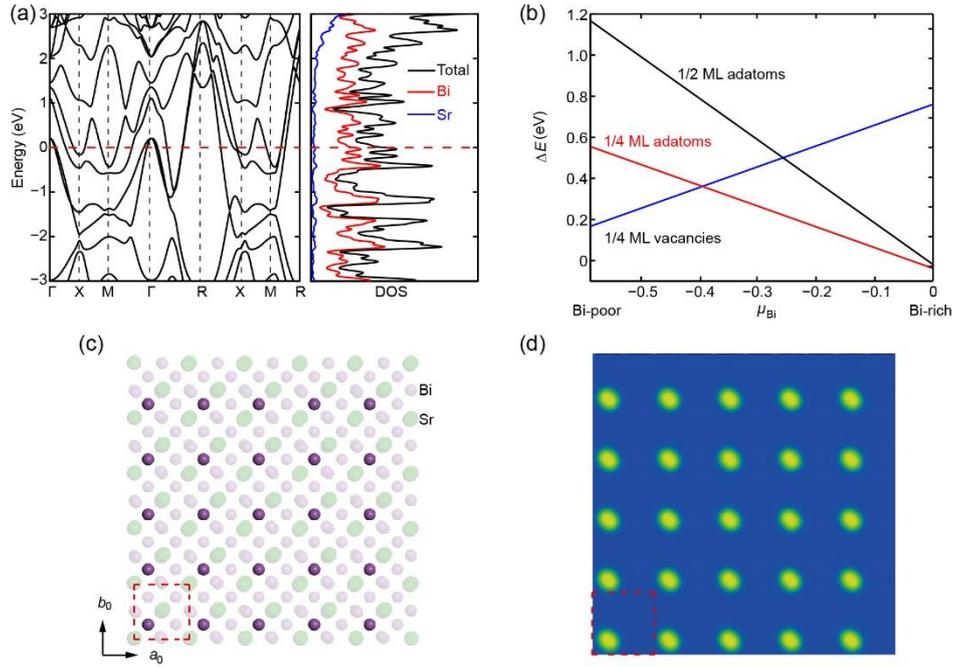

**Figure S9.** DFT calculations of SrBi$_3$. (a) Electronic band structure and DOS of bulk SrBi$_3$. (b) Surface formation energy ($\Delta E$) of $\sqrt{2} \times \sqrt{2}$ reconstructed SrBi$_3$(001) surfaces as a function of $\mu_{Bi}$, referenced to the Sr-Bi terminated surface, with $\mu_{Bi} = 0$ corresponding to bulk Bi. (c) Top view atomic structure of the Sr-Bi terminated surface with 1/4 ML Bi adatoms, and (d) its simulated STM image taken at 0.1 eV above $E_F$. A surface supercell is outlined by red dashed lines. Atoms in deep layers are shaded for clarity.

**Moiré superstructures closely matching the experimental parameters *A* and *θ*** 

| $m_0$ | 12 | 13 | 13 | 14 | 14 | 14 | 15 | 15 | 15 | 16 |
|---|---|---|---|---|---|---|---|---|---|---|
| $n_0$ | 11 | 10 | 12 | 7 | 9 | 11 | 4 | 6 | 8 | 1 |
| $m$ | 1 | 2 | 1 | 4 | 3 | 2 | 6 | 5 | 4 | 8 |
| $n$ | 12 | 12 | 13 | 11 | 12 | 13 | 10 | 11 | 12 | 9 |
| $A$(Å) | 63.6 | 64.0 | 69.1 | 61.1 | 65.0 | 69.5 | 60.2 | 63.0 | 66.4 | 62.6 |
| $\theta_{m,n}$ | 85.2 | 80.5 | 85.6 | 70.0 | 76.0 | 81.3 | 59.0 | 65.6 | 71.6 | 48.4 |
| $m_0$ | 16 | 16 | 16 | 16 | 16 | 17 | **17** | 17 | 17 | 17 |
| $n_0$ | 3 | 5 | 6 | 7 | 8 | 0 | **2** | 3 | 4 | 5 |
| $m$ | 7 | 6 | 5 | 5 | 4 | 9 | **8** | 7 | 7 | 6 |
| $n$ | 10 | 11 | 12 | 12 | 13 | 9 | **10** | 11 | 11 | 12 |
| $A$(Å) | 63.5 | 65.5 | 66.7 | 68.2 | 69.9 | 66.4 | **66.8** | 67.4 | 68.2 | 69.2 |
| $\theta_{m,n}$ | 55.0 | 61.4 | 67.4 | 67.4 | 72.9 | 45.0 | **51.3** | 57.5 | 57.5 | 63.4 |

**Table S1**. Moiré superstructures generated by use of the experimentally closer parameters of Moiré size *A* and intersection angle *θ*. The intersection angle *θ* between the lattices of $BaBi_3$ overlayer and underlying $SrTiO_3$ substrate is constrained within the interval of 40° and 50°. The black box indicates the Moiré superstructure that meets all experimental constraints.

| $H$ (T) | $N$ | $a_\Delta$ (nm) | $R_1$ (nm) | $\sigma$ (nm) | $\zeta$ (nm) |
|---|---|---|---|---|---|
| 0.2 | 22 | 109.3 | 109.7 | 17.9 | 213.9 |
| 0.4 | 48 | 77.3 | 76.0 | 11.2 | 191.4 |
| 0.5 | 21 | 69.1 | 67.4 | 9.5 | 189.1 |

**Table S2.** Best RDF fitting parameters of vortex distance $d_{ij}$ at varying field *H*. Note that the measured radius of the first coordination shell $R_1$ is compared with $a_\Delta$, despite increasing deviations with field *H*.

| $d$ (UC) | Isotropic gap | | | Anisotropic gap | | | | Fluctuating model | | |
|---|---|---|---|---|---|---|---|---|---|---|
| | $\Delta$ (meV) | $\Gamma$ (meV) | $\chi^2$ | $\Delta_1$ (meV) | $\Delta_2$ (meV) | $\Gamma$ (meV) | $\chi^2$ | $\Delta_0$ (meV) | $v_F q_0$ (meV) | $\chi^2$ |
| 4 | 0.829 | 0.220 | 0.94 | 0.674 | 0.319 | 0.122 | 0.31 | 1.045 | 0.148 | 0.09 |
| 5 | 0.897 | 0.201 | 1.08 | 0.782 | 0.301 | 0.104 | 0.30 | 1.044 | 0.087 | 0.13 |
| 7 | 0.960 | 0.204 | 1.40 | 0.849 | 0.327 | 0.089 | 0.36 | 1.056 | 0.074 | 0.32 |
| 8 | 0.956 | 0.196 | 1.27 | 0.853 | 0.328 | 0.074 | 0.48 | 1.016 | 0.055 | 0.33 |
| 11 | 1.002 | 0.178 | 1.65 | 0.927 | 0.316 | 0.056 | 0.68 | 1.029 | 0.032 | 0.66 |
| 34 | 1.059 | 0.151 | 1.99 | 1.008 | 0.287 | 0.051 | 0.70 | 0.997 | 0.013 | 0.97 |

**Table S3.** Comparison between various fit modes, i.e. superconducting fluctuation model, Dynes model with isotropic and anisotropic *s*-wave gap functions. For the thin BaBi$_3$ films, the fluctuation model gives the best description of experimental tunneling d$I$/d$V$ data.

| | Bi vacancy phase | | Bi adatom phase | |
|---|---|---|---|---|
| | homo | inhomo | homo | inhomo |
| BaBi$_3$ | 0 | 78 | 181 | 0 |
| SrBi$_3$ | 38 | 0 | 0 | 351 |

**Table S4.** Relative stability of homogeneous and inhomogeneous distribution of surface defects on Ba(Sr)Bi$_3$(001) surface. The calculated total energies (unit: meV) are listed for a homogeneous (homo) and an inhomogeneous (inhomo) distributions of 1/4 ML Bi vacancies and adatoms schematically displayed in Supplementary Fig. 7. Energy of the relatively more stable configuration is set zero.

**Supplemental References:**

[1] A. Tkatchenko, Phys. Rev. B **75**, 235411 (2007).

[2] H. I. Li, K. J. Franke, J. I. Pascual, L. W. Bruch, and R. D. Diehl, Phys. Rev. B **80**, 085415 (2007).

[3] N. N. Zhuravlev, and V. P. Melik-Adamyan, Kristallografiya **6**, 121-124 (1961).

[4] D. F. Shao, X. Luo, W. J. Lu, L. Hu, X. D. Zhu, W. H. Song, X. B. Zhu, and Y. P. Sun, Sci. Rep. **6**, 21484 (2016).

[5] R. Jha, M. A. Avila, and R. A. Ribeiro, Supercond. Sci. Tech. **30**, 025015 (2017).